\documentclass[10pt,conference,letterpaper]{IEEEtran}
\usepackage{graphicx}
\usepackage{amssymb}
\usepackage{epstopdf}
\title{Dynamic Approaches to In-Network Aggregation}

\author{%
{Oliver Kennedy{\small $~^{1}$}, Christoph Koch{\small $~^{2}$}, Al Demers{\small $~^{3}$} }%
\vspace{1.6mm}\\
\fontsize{10}{10}\selectfont\itshape
Dept of Computer Science, Cornell University\\
4130 Upson Hall\\
Ithaca, NY 14853-7501, USA\\
\fontsize{9}{9}\selectfont\ttfamily\upshape
$~^{1}$okennedy@cs.cornell.edu\\
$~^{2}$koch@cs.cornell.edu\\
$~^{3}$ademers@cs.cornell.edu
}

\date{}                                           

\begin{document}
\maketitle

\begin{abstract}
Collaboration between small-scale wireless devices hinges on their ability to infer properties shared across multiple nearby nodes.  Wireless-enabled mobile devices in particular create a highly dynamic environment not conducive to distributed reasoning about such global properties.  This paper addresses a specific instance of this problem: distributed aggregation.  We present extensions to existing unstructured aggregation protocols that enable estimation of count, sum, and average aggregates in highly dynamic environments.  With the modified protocols, devices with only limited connectivity can maintain estimates of the aggregate, despite \textit{unexpected} peer departures and arrivals.  Our analysis of these aggregate maintenance extensions demonstrates their effectiveness in unstructured environments despite high levels of node mobility.
\end{abstract}

\section{Introduction}
\label{sec:intro}


Flash memory and wireless connectivity are inexpensive. Small, portable devices such as cellphones, pdas, media players and GPS units are now capable of storing large amounts of data, carrying it wherever the owner goes, and wirelessly communicating that data to other devices in the immediate area.  Often, data on one device is also relevant to other devices in the same physical area.  For example, a gps unit can monitor car-mounted sensors to detect hazards such as slippery roads or high traffic.  Nearby units can apply this information to route around the hazards.  The ephemeral peering relationships formed between nearby devices can be exploited to spread this data.

Data transmission is limited by the transience and limited capacity of these wireless links.  However, when the desired value is an aggregate (eg, sum, average, etc..) over the data stored on the devices and not the data itself, this limitation may be mitigated.  By decomposing the computation of the aggregate across all the nodes on the network, nodes need only transmit intermediate values rather than entire datasets.  As an added benefit, the decreased bandwidth usage also reduces the device's power requirements.

In-network aggregation has been studied extensively, but typically either relies on the ability to efficiently impose structure, such as a spanning tree on the network, or produces inaccurate results if hosts leave the system during computation.  Unstructured gossip-based data aggregation protocols, such as those referenced in this paper rely on a static participant set, or at least the ability to detect departure.  In a network with an underlying routing infrastructure, this is a reasonable assumption.  However, in mobile wireless networks, peer-to-peer link failure due to devices departing the area of interest (or failing silently) is locally indistinguishable from link failure due to devices moving within the area of interest.

Consider a proximity-aware social networking application.  Participants export information about themselves and their recent activities.  For example, media players commonly allow users to rate songs stored on the device; a wireless-equipped media player can export this information to nearby devices.  Mobile devices can use information about popular songs to match people with similar taste in music.  Stationary devices (eg: in a store or bar) can use song popularity to select the optimal ambient music for their current clientele.  Similarly, a more general social networking application could provide statistics about a given area, steering users towards areas populated by those with similar interests.  The values of interest to such a system are aggregates of the data (avg song rating, most popular song, etc..), rather than the data points themselves (each user's song list).  

This paper's contribution is twofold.  First, we address the notion of in-network aggregation in dynamic environments by defining the \textit{dynamic} distributed aggregation class of protocols.  These protocols eschew two core assumptions made by static distributed aggregation protocols: 1) That there exists sufficient connectivity for hosts to establish a routing infrastructure, and 2) That participants in the computation do not fail silently.  Protocols in this class allow hosts to maintain running aggregates, even in the face of a rapidly changing network.  

Second, we demonstrate how dynamic distributed aggregation protocols may be instantiated by presenting dynamic extensions to existing static distributed averaging, counting, and summation protocols \cite{Kempe-Gossip,Considine}.  The extended protocols share the benefits of their base protocols, and additionally feature resilience to failures, departures, and other forms of network restructuring.  By allowing a small amount of error in the aggregate estimate, these protocols are able to adapt their behavior to changing network configurations.

The rest of this paper is organized as follows.  In Section 2, we provide an overview of existing static distributed aggregation protocols and introduce the idea of dynamic distributed aggregation protocols.  In Sections 3 and 4 we build on existing gossip-based aggregation protocols to create dynamic average, count, and sum aggregation protocols.  Section 5 provides an overview of our experimental results, comparing the behavior of the dynamic protocols to their static counterparts.  In Section 6 we provide a brief overview of related work, and conclude with Section 7.

\section{Distributed Aggregation}
\label{sec:aggregation}

Aggregates are representative terms computed over large datasets.  Examples of aggregates include the sum, count, average, and standard deviation. Such values are useful in analyzing the data, making decisions based on the data, or optimizing-data processing tasks.  Because of their relative size, it is often preferable to communicate aggregate values rather than entire data sets.  

This is particularly true if the data set is spread across an entire network. Collecting the entire data set at every host on the network is inefficient, or even impossible due to storage and bandwidth constraints.  Because each host must receive every value at least once, this requires linear storage space at each node and geometric network traffic.  Fortunately, many aggregate computations are decomposable, so hosts can compute such aggregates with intermediate values rather than entire data sets.  

Work on limiting the aggregate computation traffic falls into two general categories.  Overlay protocols impose structure, such as spanning trees onto hosts and use this structure to recursively compute the aggregate.  Unstructured protocols disseminate state via gossip while ensuring convergence to the aggregate via network-wide invariants.

\paragraph{Overlay protocols} The most common implementations of overlay aggregation protocols involve the construction of a spanning tree over their participant graph.  Though numerous variants exist\ \cite{Madden,Pinhero,Heinzelman}, the fundamental methodology remains the same.  As soon as a leader host registers its interest in the aggregate value, hosts arrange themselves into a communication infrastructure.  

To compute the aggregate, hosts send their values across the infrastructure towards the requester.  To minimize bandwidth use, each host aggregates its own value with all values it receives.  Thus, rather than forwarding all received values, the host need only forward the aggregate and any weight parameters necessary for downstream aggregation.  Once the value is computed by the leader, the overlay can disseminate the aggregate value if necessary.

\paragraph{Unstructured protocols} While overlay protocols are extremely efficient, there is an implicit reliance on network structure and high connectivity.  In the presence of low connectivity or high host mobility, maintaining this structure is impractical.  Unstructured aggregation protocols employ gossip-style\ \cite{Demers} techniques to gain resilience.

\textit{Gossip}, or epidemic protocols are a general class of protocol wherein each participant periodically communicates with a small number (typically one) of random peers.  During these interactions, peers forward a subset of their local state.  As more interactions occur, the systemwide state converges to the target value.  This design pattern typically provides logarithmic convergence times (in the number of participants) using a loglinear number of messages.  More importantly, gossip protocols are highly resilient to changes in network structure.  

As we show in subsequent sections, unstructured aggregation protocols build on the resilience of gossip protocols.  During each gossip exchange, a host forwards its current view of the network-wide aggregate.  Upon receiving it, the receiving peer updates its own estimate by merging the received view with its own.  The stability of the value on which the protocol converges is maintained by choosing a merge operator that alters each node's local view, while maintaining an invariant (theoretical) systemwide view.  For example, the bitwise OR operator is duplicate-insensitive and may be used in this way.  Similarly, zero sum-exchanges maintain a constant systemwide sum.

Informally, every exchange drives each communicating host's state closer to the aggregate independently, while any estimate that can be derived from their combined state remains unchanged.  By extension, the estimate derivable across the entire network also remains unchanged.

\subsection{Distributed Averaging}
One example of a distributed aggregation protocol is Kempe et al.'s Push-Sum\ \cite{Kempe-Gossip} averaging\footnote{Push-Sum is also capable of computing the network-wide sum, though this requires knowledge of network size.  We return to this in Section \ref{sec:average}} protocol.  One iteration of this algorithm is shown in Figure \ref{fig:pushsum}.  Every host participating in push-sum maintains two values: a weight $w_{i,t}$ and a sum $v_{i,t}$, respectively initialized to 1 and the host's initial value.  The vector of these these two values is referred to as the host's mass.  At every iteration of the protocol, each host selects a random participant to communicate with.  It sends half of its current mass to the other participant, and the other half to itself.  It then sets its own new mass for the next protocol iteration to the sum of all received mass.  After convergence, the systemwide average can then be estimated at any host as:

$$Average = \frac{v_{i,t}}{w_{i,t}}$$

\begin{figure}
\begin{enumerate}
\item Select a random peer $P$
\item Send identical messages $\left<\frac{w_{t}}{2},\frac{v_{t}}{2}\right>$ to both $P$ and Self
\item Collect all messages $\left<\hat{w}_i,\hat{v}_i\right>$ received during iteration $t$
\item Updated weight $w_{t+1} := \sum_i \hat{w}_i$
\item Updated value $v_{t+1} := \sum_i \hat{v}_i$
\item $Estimate_{t+1} := \frac{v_{t+1}}{w_{t+1}}$
\end{enumerate}
\caption{\textbf{Kempe et al.'s Push-Sum algorithm}}
\label{fig:pushsum}
\end{figure}

Intuitively, the flow of mass away from each host is proportional to the mass at the host, while the flow of mass towards a host is proportional to the total mass in the system.  These flows do not reach equilibrium at a given host until the value at the host has converged to the systemwide average.  The weight provides a normalization factor to compensate for differences between mass sent and received during any given gossip iteration.

The convergence of this protocol is closely tied to the zero-sum exchanges it employs.  Because mass is neither created nor destroyed during these exchanges (a property referred to as conservation of mass), the value on which the system converges also remains constant.  A node sends half of its mass away at every iteration, so old errors are reduced by a factor of 2 at every iteration.  Because the expectation of the in-flow value remains constant, the expected error decreases exponentially at every iteration.

However, unexpected node departures result in mass leaving the system.  Nodes that exit with a different amount of mass than they entered with cause the mass to diverge from its expected value.  Over time, this increases the systemwide error, especially when node failures are correlated their local values.  We address this in Section \ref{sec:average}, where we show how introducing small amounts of local error results in a system that reverts to stability, despite silent failures.

\subsection{Distributed Counting}
Counting sketches were originally introduced by Flajolet and Martin\ \cite{Flajolet} as a means of estimating the number of unique elements in a large data set.  As demonstrated by Considine et al.\ \cite{Considine}, they are also applicable to in-network counting and summation.  Counting sketches employ the integer function $\rho$
$$\rho(i) : \mathbb{Z} \rightarrow [0, L]$$
where $\rho(i)$ is obtained deterministically with the probability distribution
$$P[\rho(i) = k] = \frac{1}{2^{k+1}}$$

While other definitions exist, $\rho$ is defined canonically as the index of the first nonzero bit of the $L$-bit cryptographic hash of $i$, or the value $L$ in the case that the hash contains only zeroes. $\rho$ assigns every object $i$ a non-unique $L$-bit bitstring
$$B_i = 2^{\rho(i)}$$

The sketch $A$ is obtained as the bitwise OR of all $B_i$.  This sketch can be divided into three regions: The low order bits will with high probability be set to one, the high order bits will with high probability be set to zero, and a region in the middle will contain both.  The function $R(A)$, equal to the size of the segment of contiguous ones is related (as a consequence of the distribution of $\rho(i)$) to the number of unique objects $n$ in the set by
$$R(A) \approx \log_2(\varphi n)$$
where $\varphi \approx 0.77351$. This sketch can be used to estimate $n$.
$$n \approx \varphi2^{R(A)}$$

Unfortunately, the variance of this estimate is high.  To address this, Flajolet and Martin present a stochastic averaging process\ \cite{Flajolet}.  Each object is deterministically sorted into one of $m$ bins $B_{i}[n]$, and one sketch is generated for each bin.  Adding bins reduces variance at the cost of increasing storage requirements.  Processing time is not significantly impacted, as each object is still counted only once.

As Consadine et al\ \cite{Considine} note, the sketch has two properties that make it useful for distributed counting protocols. First, the computation continues unaltered if the input set is split into $n$ independent parts, each generating bitstring $A_n$; $A$ is the bitwise OR of all $A_n$. Second, $A$ is insensitive (by design) to duplication between $A_n$; Intermediate stages of a computation of $A$ that both contain $A_n$ may be safely combined.

\begin{figure}
\begin{enumerate}
\item Select a random peer $P$
\item Send all local bit bins $B_{t}[n]$ to both $P$ and Self
\item Collect all bins $\hat{B}_i[n]$ received at iteration $t$
\item Updated bin $n$, $B_{t+1}[n] :=$ the bitwise OR $\forall i$ of $\hat{B}_i[n]$
\item $Estimate_{t+1} := |B|\cdot \varphi\cdot 2^{Avg_n(R(B_{t+1}[n]))}$
\end{enumerate}
\caption{\textbf{Consadine et al.'s Sketch-Count algorithm}}
\label{fig:sketchcount}
\end{figure}

As a consequence, hosts can exchange bitstrings (ORing the received bitstring with their own) without affecting the systemwide estimate of the sum.  The number of hosts in the network can be computed by storing one ``object" at each host.  Sums can be computed by storing $n$ objects at a host, where $n$ is that host's value.  This protocol is summarized in Figure \ref{fig:sketchcount}.

Each new host that enters the system brings with it an additional identifier, and an additional chance for a high-order bit to be set.  Consequently, unless hosts remove their contribution to the systemwide bit vector before departing, the estimate increases monotonically.  However, even if a host can inform the network of its departure, it cannot independently determine whether it is the only host sourcing that particular bit.  We address this in Section \ref{sec:count}, where we show that a network size-agnostic timeout exists for each bit and can be used by participating hosts to determine when a bit is no longer being sourced.

\subsection{Dynamic Aggregation}
Distributed aggregation protocols as described above are designed to be executed in a query/response manner.  Overlay protocols are reliant on a leader to initiate construction of the overlay and evaluation of the aggregate.  Thus, the leader host initiates the computation, and the response is returned to it.

By comparison, unstructured protocols require no leader and are resilient to changes in network structure.  However, changes in the network's composition cause the aggregate to change.  Host joins can be handled transparently; the random nature of gossip communication requires that protocols cope with hosts disconnected by pure chance.  However, this same feature makes it difficult to detect when a host has been truly removed from the computation.  This is especially evident in mobile wireless environments, where host departures are indistinguishable from hosts moving within network.  

As a host is removed from the network, its impact on the systemwide aggregate must also be removed.  By design, each host's initial value is stored only at the host itself.  Where it is infeasible for the host to gracefully depart the network (i.e., by performing a sign-off protocol), an error is introduced into the computation.  Limiting the protocol to query/response limits the number of rounds spent computing the aggregate.  Damage is limited to host failures that occur during this period, as the protocol begins anew with every query.

Without the ability to detect host departures or to impose structure on the network, protocols must resort to other means of current values.  In this paper, we contribute two instances of \textit{dynamic} distributed aggregation protocols.  Unlike the previously described \textit{static} distributed aggregation protocols, these new protocols provide each host with a \textit{continual} estimate of the current systemwide aggregate.

The simplest form of dynamic aggregation is the use of epochs with an existing aggregation protocol.  At periodic intervals, the network resets the aggregation protocol to its initial state and begins anew.  Without resorting to a leader, this may be done via weak clock synchronization by annotating each message with a periodically incremented epoch counter.  However, in the case of mobile wireless hosts, hosts traveling from one clique of hosts to another will encounter variance in epoch number.  Thus node mobility may result in disruptions in aggregate computation while the destination clique settles on a new epoch number.

While this technique is highly generalizable from an algorithmic standpoint, the optimal epoch length is closely tied to the size of the network.  Gossip convergence time is a function of the size and structure of the network, so if the epoch length is too small then the protocol may reset before it converges.  However, overestimating convergence time results in unnecessarily coarse results.  Ironically, computing the size of the network is another aggregate computation.  This cyclic dependence reduces the effectiveness of epoch-based protocols in dynamic environments.

\section{Distributed Averaging}
\label{sec:average}
In Push-Sum, a host's mass (defined as the sum $v_{i,t}$ and weight $w_{i,t}$ values stored at the host) remains constant throughout the computation.  The protocol relies on this ''conservation of mass`` property.  Silent host failures introduce error, both as a consequence of the mass removed when the host fails and due to the resulting change in system average.  The error introduced is especially high if there exists a correlation between host failures and host values.

Our first protocol contribution addresses these issues by introducing a controlled local error at each host that adjusts local equilibrium towards the host's initial value.  The \textit{Push-Sum-Revert} protocol takes a reversion constant $\lambda$ as a systemwide input.  After every gossip iteration, the protocol updates each host's mass. 
$$w_{i,t+1} :=\lambda + (1-\lambda)\cdot \sum_r\hat{w}_{r,t}$$
$$v_{i,t+1} :=\lambda \cdot v_{i,0} + (1-\lambda)\cdot \sum_r\hat{v}_{r,t}$$
Where $\hat{w}$ and $\hat{v}$ represent the mass received during the prior iteration.  One iteration of this protocol is summarized in Figure \ref{fig:pushsumrevert}.

\begin{figure}
\begin{enumerate}
\item Select a random peer $P$
\item Send identical messages $\left<\frac{(1-\lambda)w_{t} + \lambda}{2},\frac{(1-\lambda)v_{t} +\lambda v_{0}}{2}\right>$ to both $P$ and Self
\item Collect all messages $\left<\hat{w}_i,\hat{v}_i\right>$ received during iteration $t$
\item Updated weight $w_{t+1} := \sum_i \hat{w}_i$
\item Updated value $v_{t+1} := \sum_i \hat{v}_i$
\item $Estimate_{t+1} := \frac{v_{t+1}}{w_{t+1}}$
\end{enumerate}
\caption{\textbf{The Push-Sum-Revert protocol}}
\label{fig:pushsumrevert}
\end{figure}

Push-Sum-Revert can be expressed as the composition of the classic Push-Sum protocol and a Revert step that causes the mass at each host to decay towards its initial value.  We now show that the Revert step does not change the mass in the system.
$$\sum_i revert_v(v_{i,t}) = \sum_i\left[ (1 - \lambda) v_{i,t} + \lambda v_{i,0}\right]$$
$$= \left(\sum_i v_{i,t}\right) - \lambda \left(\sum_i v_{i,t}\right) + \lambda \left(\sum_i v_{i,0}\right)$$
At time $0$, the last two terms are equal and drop out.  The remaining term, $\sum_i v_{i,t}$ is identical to the input, so revert obeys conservation of mass up to this point.  If conservation of mass is obeyed up to time $t$ and nodes neither enter nor leave the system, then $\left(\sum_i v_{i,t}\right) = \left(\sum_i v_{i,0}\right)$ and the last two terms drop out.  
$$\sum_i revert_v(v_{i,t}) = \sum_i v_{i,t}$$
The same line of reasoning can be applied to $w_{i,t}$.  Thus by recursion, if the node set remains unchanged then the push-sum-revert protocol obeys conservation of mass.  

While this does not prove the convergence of the variant (convergence is shown experimentally Section \ref{sec:eval}), it provides intuition as to why the protocol converges to an approximation.  Under equilibrium, the value at a host is lowered at hosts where the initial value is below the average.  For all the mass removed in this manner, an equal amount of mass is inserted by nodes with an initial value above the average.

\subsection{Improving the Estimate}

Push-Sum-Revert functions by introducing an error at each host.  While this error pushes the host (and thus the entire network) towards convergence, it also places a hard limit on the accuracy of the host's estimate.  In a nonuniform gossip environment, this introduces a local distortion in the hosts estimate of the system average.  The error is proportional to the difference between the host's initial value (as well as the initial values of nearby hosts) and the system average.  However, in a uniform gossip environment (or an equivalent setting), this same distribution of errors can be exploited to obtain a more accurate estimate.

Counterintuitively, increased accuracy over time is achieved by increasing the variance in a given host's estimate over any given gossip cycle.  If a host exports its entire mass at any given gossip iteration, it is forced to rely exclusively on imported mass to obtain an estimate.  Though the variance in the estimate at any given iteration is increased, the host's initial value no longer plays a role in the correlation between successive estimates.  The results of the $T$ most recent protocol iterations average to obtain a more accurate result.  The algorithm with the Full-Transfer optimization is summarized in Figure \ref{fig:pushsumrevertfulltransfer}.

There exists a danger that a given host will not receive any mass at a given iteration.  To mitigate this, hosts divide their mass into a number ($N$) of parcels rather than just 2.  These parcels are then distributed to independently selected peers.  Though doing so reduces reaction time, hosts can also generate their estimate based on the last $T$ iterations during which they received mass.  This approach is preferable in mostly stable systems, as it does not result in random increases in variance.

\begin{figure}
\begin{enumerate}
\item Select $N$ random peers $P_n$
\item Send identical messages $\left<\frac{(1-\lambda)w_{t} + \lambda}{N},\frac{(1-\lambda)v_{t} +\lambda v_{0}}{N}\right>$ to all $P_n$
\item Collect all messages $\left<\hat{w}_i,\hat{v}_i\right>$ received during iteration $t$
\item Updated weight $w_{t+1} := \sum_i \hat{w}_i$
\item Updated value $v_{t+1} := \sum_i \hat{v}_i$
\item $Estimate_{t+1} := \frac{\sum_{i=t-T}^{t+1}v_{i}}{\sum_{i=t-T}^{t+1}w_{i}}$
\end{enumerate}
\caption{\textbf{The Push-Sum-Revert protocol with Full-Transfer optimization}}
\label{fig:pushsumrevertfulltransfer}
\end{figure}

As a more general observation, the communication graph during any given iteration of a push gossip protocol has a high variance in host indegree.  Some hosts receive messages from a large number of peers, while others may receive no messages at all.  In this push-gossip model where hosts contact random peers and push state to them, some hosts may not receive any state at all.  As observed by Karp et al\ \cite{Karp}, after state has been pushed to half the hosts in the network, gossip favors a pull-based model where hosts seek out state stored at other hosts.  Specifically, the {\em initial} convergence time of Push-Sum is nearly halved under uniform gossip when it applies a pushpull gossip model.  Under this optimization, each host exports (or imports) half the difference between its own mass and the mass of its communications peer, rather than half of its total value.

A similar optimization reduces the time it takes Push-Sum-Revert to reconverge after host failure where host values are uniformly distributed.  The higher a host's indegree (i.e., the number of hosts it receives mass from), the more mass it receives at that iteration.  This additional mass works to counteract the reversion constant $\lambda$.  Consequently, hosts with higher indegrees can use a larger reversion constant.  Rather than adding a fixed $\lambda$ factor of its initial mass, a host adds $\frac{\lambda}{2}$ for every message it receives including the one it sends to itself.  Half of a host's mass is contained in every message, so on average a host will add $\lambda$ exactly once per iteration.  Under a uniform value distribution, the expected change in systemwide average remains 0, while reconvergence times are approximately halved.  Alternatively, a lower $\lambda$ may be used to provide equivalent convergence times with a lower error.

\section{Dynamic Counting}
\label{sec:count}
Counting sketches provide a flexible means of rapidly computing an estimate of the size of a network.  From an abstract standpoint, the protocol is based on the distribution of gossip messages, the number of which is proportional to the logarithm of the network size.  Hosts maintain a list of these messages and use the number of unique messages received to to compute the actual network size.  

In Count-Sketch, two hosts that select the same bit as their identifier generate identical messages.  This redundancy is a large component of the system's scalability, but at the same time interferes with self-healing.  Even if a host broadcasts its identifier to the network upon departing, a host that receives the departing identifier has no reliable way to independently determine if another (still active) host continues to originate the message (i.e., there is another host that uses the same bit as its identifier).  Reversion as used in Push-Sum-Revert is unsuited to this protocol because binary messages do not admit decay.  

Self-healing requires that the capacity for decay be introduced into the protocol.  Consider a network where each host has a single bit $B_i$, initially set to zero.  Before the test is run, a fraction of the hosts set their own bit to 1.  Each host also maintains an integer $N_i$ initialized to $\infty$ if $B_i = 0$ and initialized to $1$ otherwise.

At every iteration, all hosts where $B_i \neq 1$ increment their integer.  All hosts then select a random peer to gossip $N_i$ to.  Subsequently, each host updates its own integer to the lowest of its current value or any value received.  This simplified protocol has three interesting properties: 
\begin{itemize}
\item Finite values of $N_i$ are bounded by the number of gossip rounds elapsed minus 1.

\item In the absence of hosts at which $B_i = 1$, the minimum noninfinite value of $N_i$ will increase by 1 at every iteration.

\item The maximum value of $N_i$ can be bounded with high probability, as long as there exists at least one host where $B_i = 1$.
\end{itemize}

The third property is nontrivial.  Intuitively, every gossip round, all hosts where $B_i = 1$ begin to gossip a new message.  Every host maintains a record of and continues to gossip about the youngest message it has received to date.  As a consequence, the maximum value of $N_i$ is related to the expected convergence time of the network.

Over uniform gossip, this bound is proportional to the logarithm of the network size.  As observed by Demers et al\ \cite{Demers} and subsequently proven by Kempe et al\ \cite{Kempe-Spatial}, this is also approximately true of spatially distributed gossip environments.  Given hosts distributed uniformly in an n-dimensional space, logarithmic convergence times can be achieved by sending a limited number of long-distance communications (eg, via random walks, an underlying routing infrastructure, or host mobility).  Even in this case however, the bound is still contingent on a the presence of a host where $B_i = 1$.

While network size is not known in advance, the bound is also inversely proportional to the number of hosts sourcing the message.  Karp et al\ \cite{Karp} note in their analysis of gossip propagation that the number of affected hosts grows exponentially until approximately half the nodes have received the message (in push-gossip).  Conversely, from this point, each halving of the number of affected hosts increases convergence time linearly.  

Once half of the nodes are aware of the message, the fraction of unaware nodes is reduced by a constant factor at every round.  The number of rounds required for any given fraction of nodes to become aware of the message is constant in the size of the network.  Thus, 
$$\max(N_i) \propto \log(P[B_i = 1])$$ 

\subsection{Count-Sketch-Reset}

Returning to counting sketches, the probability that any given host will source bit $k$ is $\frac{1}{2^{k+1}}$.  Consequently, if each bit $k$ is replaced with an integer $N_{i,k}$, then $\max(N_{i,k})$ is linear in $k$.  The bound on $N_{i,k}$ is agnostic to the network size.  With this bound, a node can independently determine with high probability whether a particular bit is still being sourced without knowing the size of the network. 

The \textit{Count-Sketch-Reset} protocol maintains a two-dimensional array of \textit{integers} $N_{H}[n][k]$ at each host, as opposed to the two-dimensional array of \textit{bits} used in Count-Sketch.  The integers are initialized to infinity, and each host selects one pair of matrix indices as per the standard distributions used for counting sketches.
$$P[n \in [0,m)] = \frac{1}{m}$$
$$P[k \in [0,L]] = \frac{1}{2^{k+1}}$$
Where $m$ is the number of bins.  

Each host sets the integer at its selected index to zero.  Every round of gossip proceeds in two phases.  First, each host increments all values in its array (except at its selected index) by one.  Subsequently, each host selects a peer to send its array to.  The peer then examines every element of its own array, and sets the element's value to the corresponding value of the incoming array if it is lower.  In order to accelerate convergence (and consequently lower the bound on $N_{i}$), the peer can also respond by sending its own array.  The Count-Sketch-Reset protocol is summarized in Figure \ref{fig:sketchcountreset}.

\begin{figure}
\begin{enumerate}
\item Select a random peer $P$
\item Increment local integers $N_{t}'[n][k] := N_{t}[n][k] + 1$ excluding the host's selected bin/index pair.
\item Send all incremented integer bins $N_{t}'[n]$ to both $P$ and Self
\item Collect all bins $\hat{N}_i[n]$ received at iteration $t$
\item Update local bins $N_{t+1}[n][k] := min_i(\hat{N}_i[n][k])$
\item Generate the $k$th bit of bit bin $B_{t+1}[n][k] := N_{t+1}[n][k] \leq f(k)$
\item $Estimate_{t+1} := |B|\cdot \varphi\cdot 2^{Avg_n(R(B_{t+1}[n]))}$
\end{enumerate}
\caption{\textbf{The Sketch-Count-Reset protocol}}
\label{fig:sketchcountreset}
\end{figure}

\begin{figure}
\begin{center}
\includegraphics[width=3.5in]{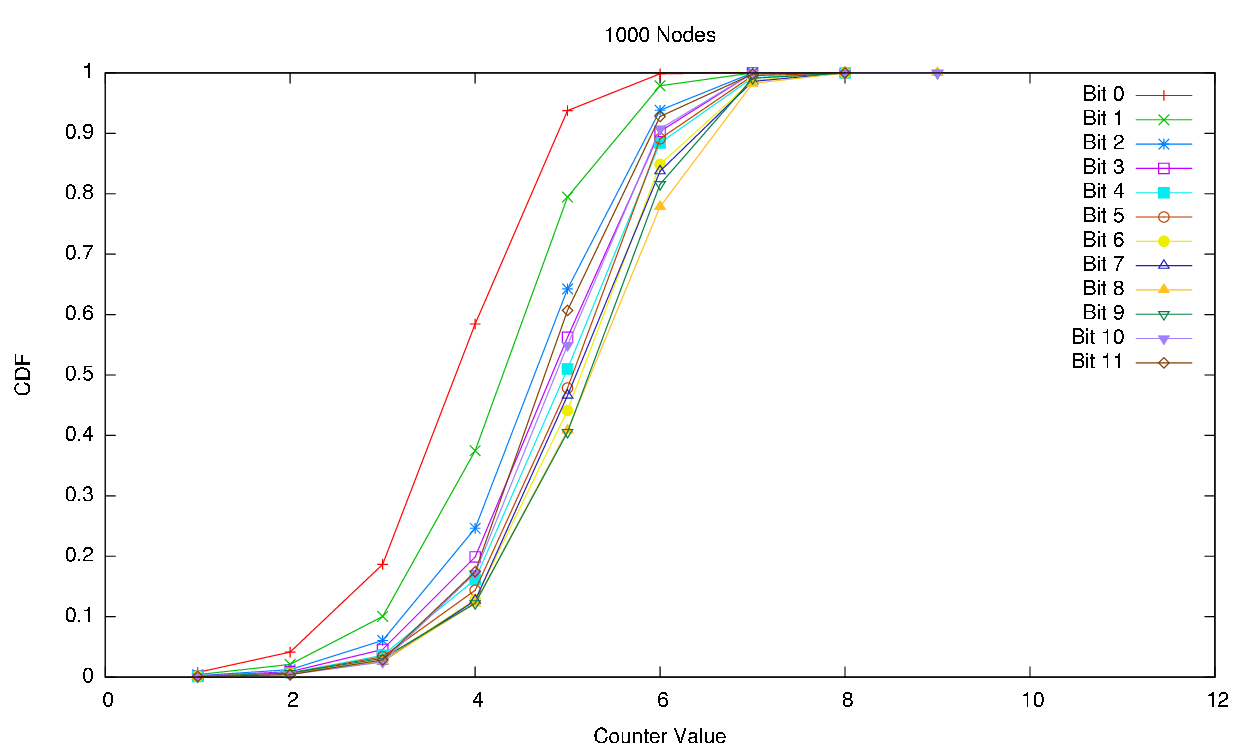}
\includegraphics[width=3.5in]{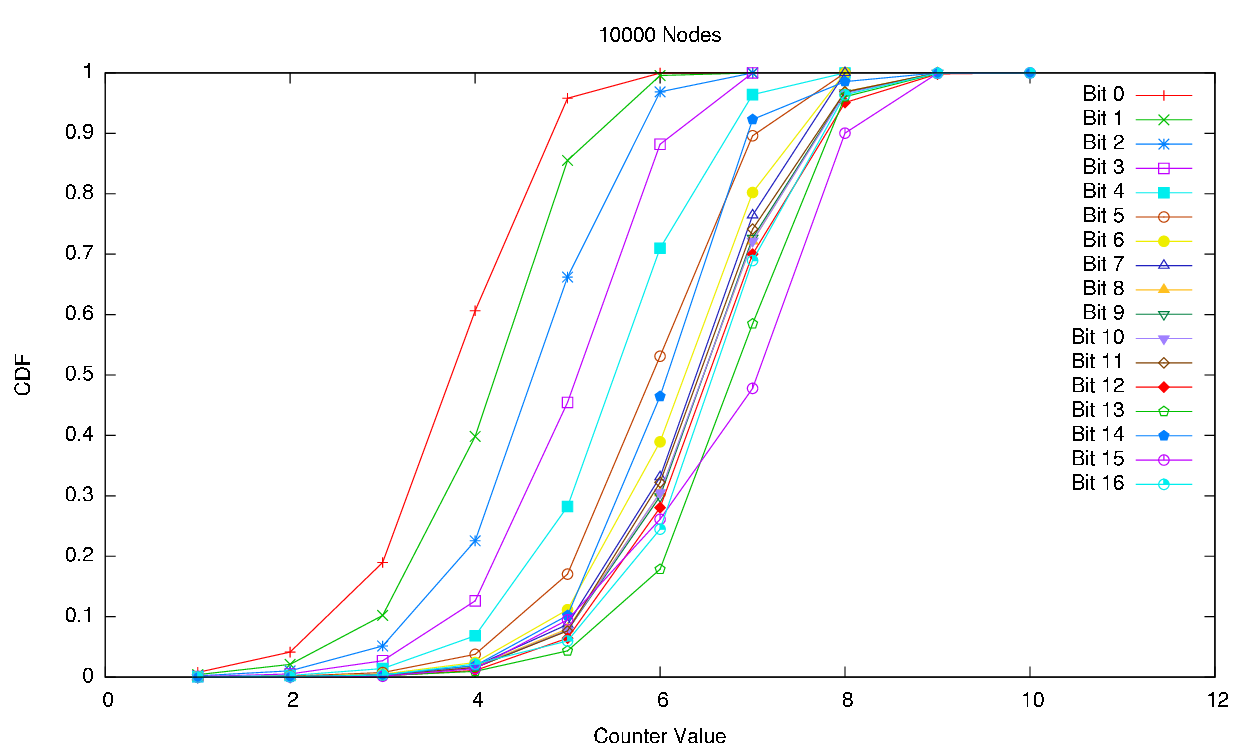}
\includegraphics[width=3.5in]{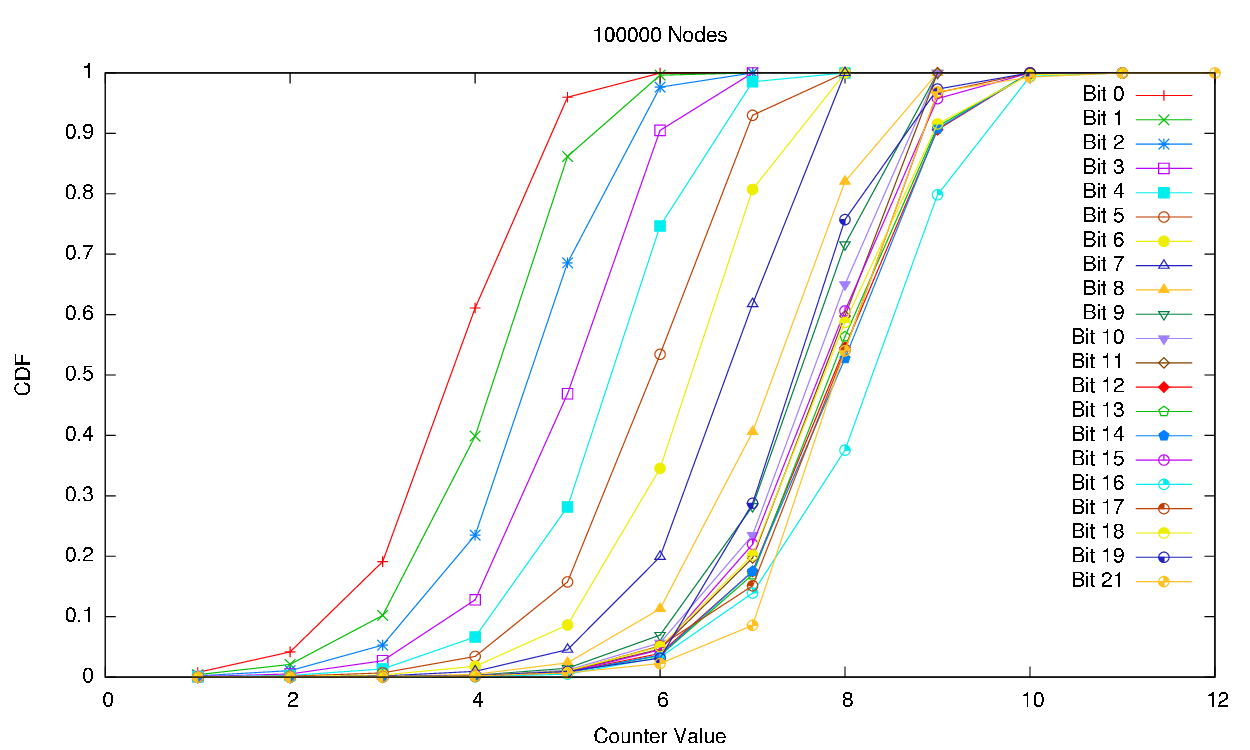}
\caption{\textbf{Bit counter distribution.} Each graph shows a CDF plot of the bit counters of a simulated fully converged network of varying host count.  Each line represents the counter for the indicated bit.  As predicted, the distribution distance of the lower order bits increases linearly.}
\label{fig:bitdistance}
\end{center}
\end{figure}

The protocol obtains an estimate of the network size by means of an expected maximum convergence time function $f(k)$.  We have derived this function experimentally based on data summarized in Figure \ref{fig:bitdistance}.  This figure shows the distribution of counter $N_{i}$ values with respect to the index $k$.  As the size of the network increases, the distribution of counter values (save for a tail at the high indices) remains constant.  The distribution of counters for index $k$ can be bounded with high probability in this uniform gossip environment by a linear function of $k$.
$$f(k) \sim 7 + \frac{k}{4}$$

Each host derives a bit array from its integer array by partitioning entries into those greater than $f(k)$ (and setting the corresponding bit to 0) and those less than or equal to $f(k)$.  The bit array is converted to the base 2 logarithm of the network size via Flajolet and Martin's $R$ function.  From this point, the estimates are averaged and the result is computed exactly as in Sketch-Count.

This cutoff is determined based on the gossip propagation rate of the network.  As shown above, in a uniform gossip environment, this cutoff is a linear function of $k$.  However, a similar bound\ \cite{Kempe-Spatial} may be achieved even in spatially distributed environments, where hosts distributed evenly in a D-dimensional grid can only communicate with adjacent nodes.  This approximation requires the use of multi-hop messages where the probability of sending a message to a node within $d$ hops is proportional to $\frac{1}{d^2}$.  After a random distance $d$ is selected by the source node, the network approximates random peer selection by performing a random walk of length $d$.  

Unlike the grid peering used in spatially distributed gossip, wireless devices can communicate with all devices in range at roughly constant cost.  For densely packed networks, a greater ``distance" can be achieved without increasing bandwidth usage.  While density may not be uniform, Push-Sum-Revert may be used to compute average node degree and the multi-hop distance may be selected accordingly.

\subsection{Summation}

As Considine et al demonstrate in their treatment of counting sketches, sketches may also be used to compute sums.  To register the value $v$, a node acquires $v$ independent identifiers.  The additional identifiers can be stored in the same bit vector, so the only space and performance overhead introduced by this process is due to the need for a bigger bit vector (the size of which scales logarithmically with the range of $v$).

Though the multiple insertions technique has good scaling properties, it is possible to achieve even higher space and bandwidth efficiency at the cost of some precision.  Count-Sketch-Reset can be used to compute the number of nodes in the network, while Push-Sum-Revert computes the average value in the network.  The two values multiplied together are an estimate of the network-wide sum.  Though errors introduced by both protocols multiply, Push-Sum-Revert requires several orders of magnitude less bandwidth and storage space than Count-Sketch-Reset.  Since the cost of the latter may be amortized over multiple summations, this \textit{Invert-Average} protocol is significantly less expensive than the multiple insertion technique.  The protocol is summarized in Figure \ref{fig:invertaverage}.

\begin{figure}
\begin{enumerate}
\item Compute $netsize_t :=$ Count-Sketch-Reset()\\
\item For each desired value $v$\\
compute $A_{v,t} :=$ Push-Sum-Revert($v$)
\item $Estimate_{v,t} := \frac{A_{v,t}}{netsize_t}$
\end{enumerate}
\caption{\textbf{The Invert-Average protocol}}
\label{fig:invertaverage}
\end{figure}

\section{Evaluation}
\label{sec:eval}
Gossip {\em protocols} are distinct from gossip {\em environments}.  While the former defines the exchange performed by participating hosts, the latter defines how pairs of hosts are selected to perform an exchange.  We have implemented our protocols in a simulator capable of simulating three separate gossip environments.  

Our simulator employs a common simplification used to analyze gossip protocols: simulation in rounds, or iterations.  At every iteration, each host performs the protocol's exchange with one peer, selected as per the environment.  As a consequence, every push/pull iteration requires a minimum of $2n$ messages, where $n$ is the number of participating hosts.  Except as noted below, when hosts are required to have values, the values are selected uniformly in the range $[0,100)$.  Errors are presented in aggregate as the standard deviation from the correct value.

The uniform gossip environment consists of 100,000 hosts with full connectivity.  Idealized models of this form are commonly employed in the analysis of gossip protocols.  Despite their simplicity, they can serve as an effective estimator of a gossip protocol's real-world performance.  Tests in the uniform gossip environment allow nodes 20 gossip rounds to converge, before failing half of the participating nodes.

The example of a media player-based social networking application introduced earlier in the paper forms the basis for the second environment.  We simulate the behavior of devices in this environment using the CRAWDAD Cambridge/Haggle datasets\ \cite{cambridge-haggle-2006-09-15}.  These three datasets trace changes in the adjacency matrix of wireless-equipped devices as they are carried by people going about their daily lives in the first two traces and attending a conference in the third.  Each trace encompasses between 9 and 41 devices.  In the simulation, each device is treated as a participating device.  Devices perform one round of gossip every thirty seconds of simulated time, and hosts are restricted to communicating with hosts in wireless range.  

At every gossip round, a host is assigned to a group of all ``nearby" hosts.  Because the traces do not include location information, two hosts are ``nearby" if there exists a path from one to the other over the union of all edges that have existed in the last 10 minutes.  A host's error is reported relative to the aggregate of its group.

\begin{figure}
\begin{center}
\includegraphics[width=3.5in]{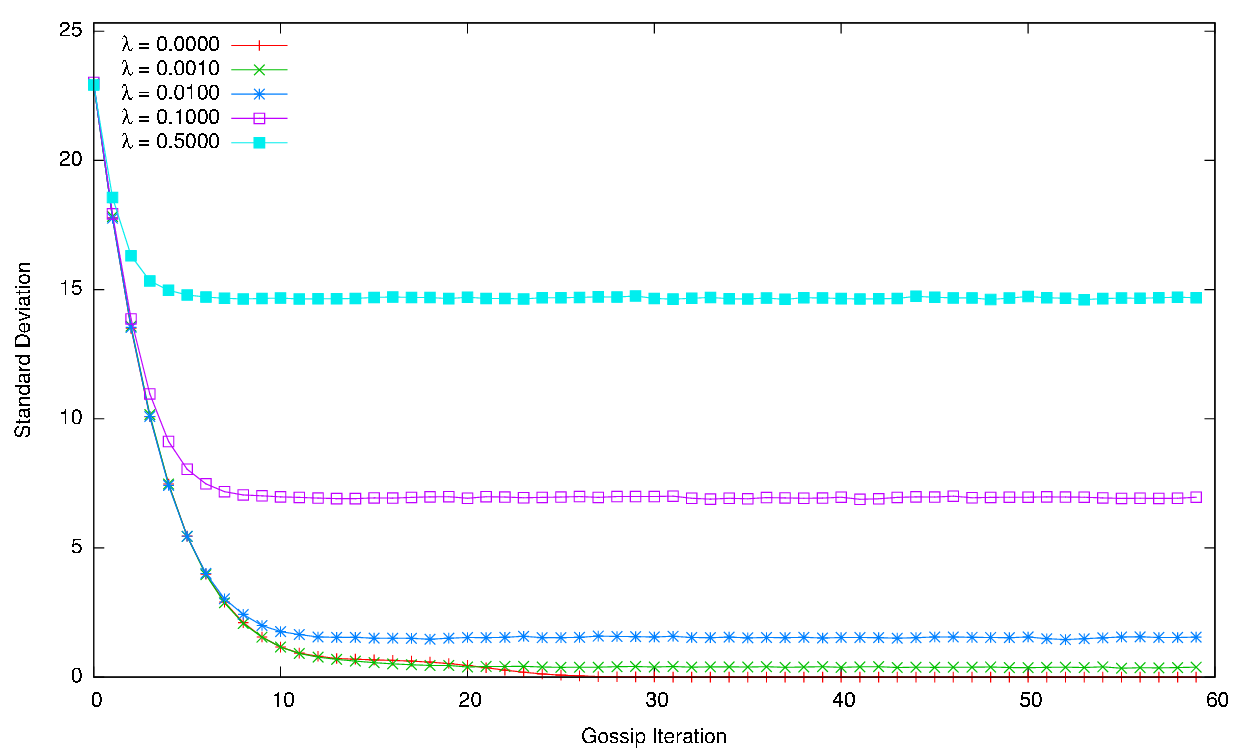}
\caption{\textbf{Accuracy of dynamic averaging under uncorrelated failures.} The standard deviation from the correct average as a function of gossip iterations.  100000 hosts were each assigned a uniformly distributed random value.  At every iteration, all hosts performed a push/pull exchange with one randomly selected peer.  After 20 iterations, 50000 random hosts were removed.  Each line represents a different reversion constant.}
\label{fig:average:uncorrelated}
\end{center}
\end{figure}

\begin{figure}
\begin{center}
\includegraphics[width=3in]{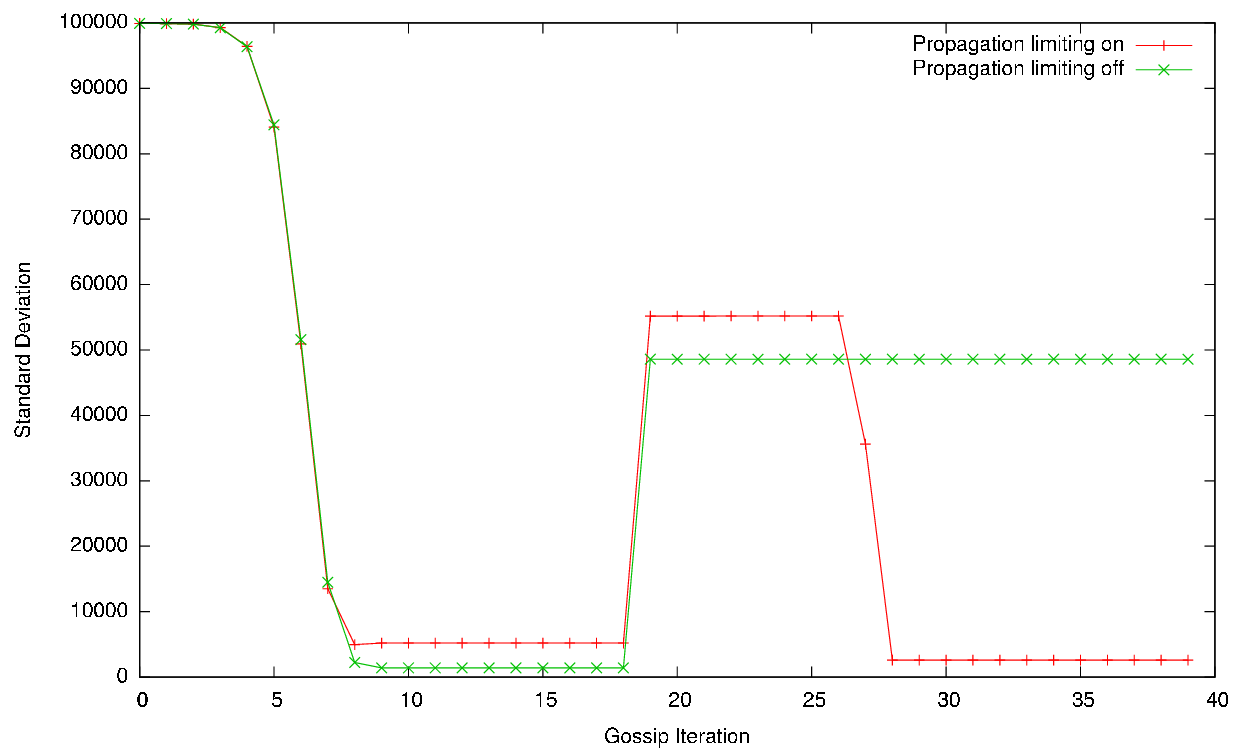}
\caption{\textbf{Accuracy of the dynamic counting under failure.} The standard deviation from the correct sum as a function of gossip iterations.  100000 hosts were each assigned a value of 1.  After 20 rounds of gossip, half the hosts were removed.  The graph shows two lines, one showing the naive sketch counting and one showing the effect of limiting the propagation of each bit $k$ to a distance of $7+\frac{k}{4}$.}
\label{fig:sum}
\end{center}
\end{figure}

\begin{figure*}
\begin{center}
\begin{tabular}{cc}
\includegraphics[width=3.5in]{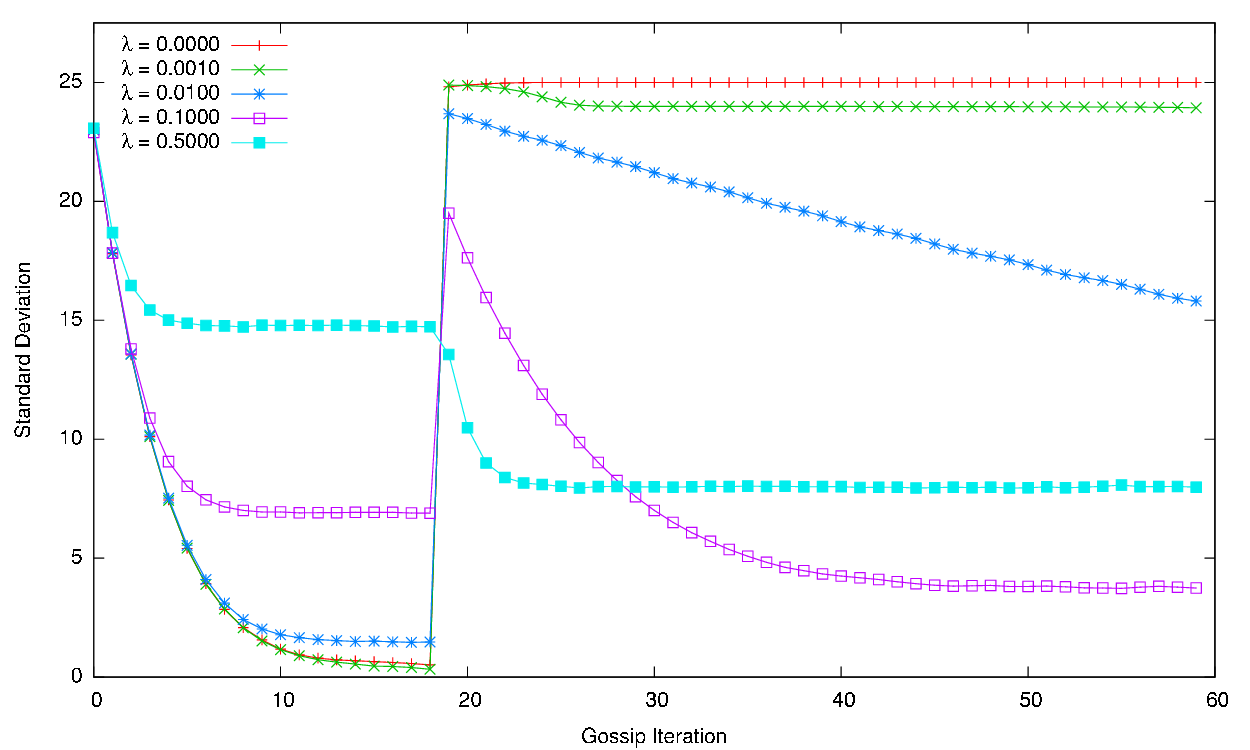} & 
\includegraphics[width=3.5in]{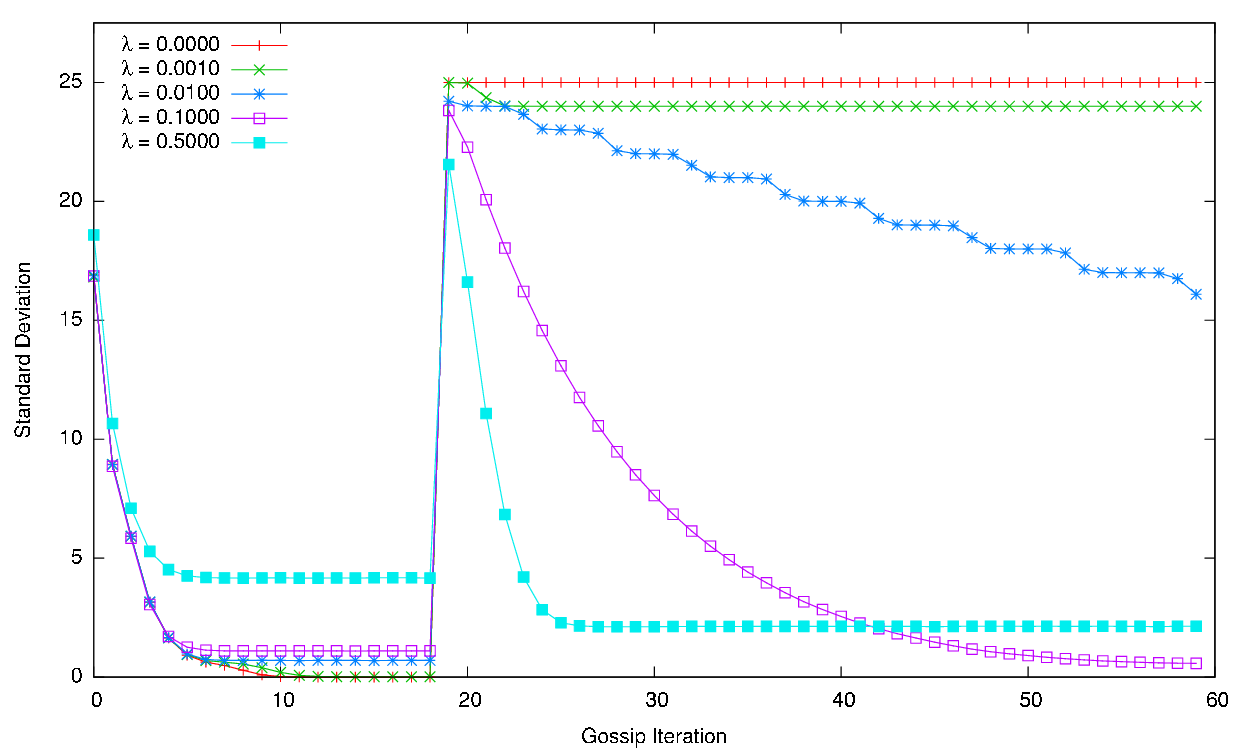}
\\(a) & (b)\end{tabular}

\caption{\textbf{Accuracy of dynamic averaging under correlated failures.} The standard deviation from the correct average as a function of gossip iterations.  The error is normalized based on the current average.  Each line represents a different reversion constant.  A reversion constant of 0 is equivalent to the unmodified (static) protocol.  Graph (a) shows the results of the basic algorithm.  Graph (b) shows the results of the Full-Transfer optimization.  }
\label{fig:average:correlated}
\end{center}
\end{figure*}

\begin{figure*}
\begin{center}
\begin{tabular}{cc}
\includegraphics[width=3.5in]{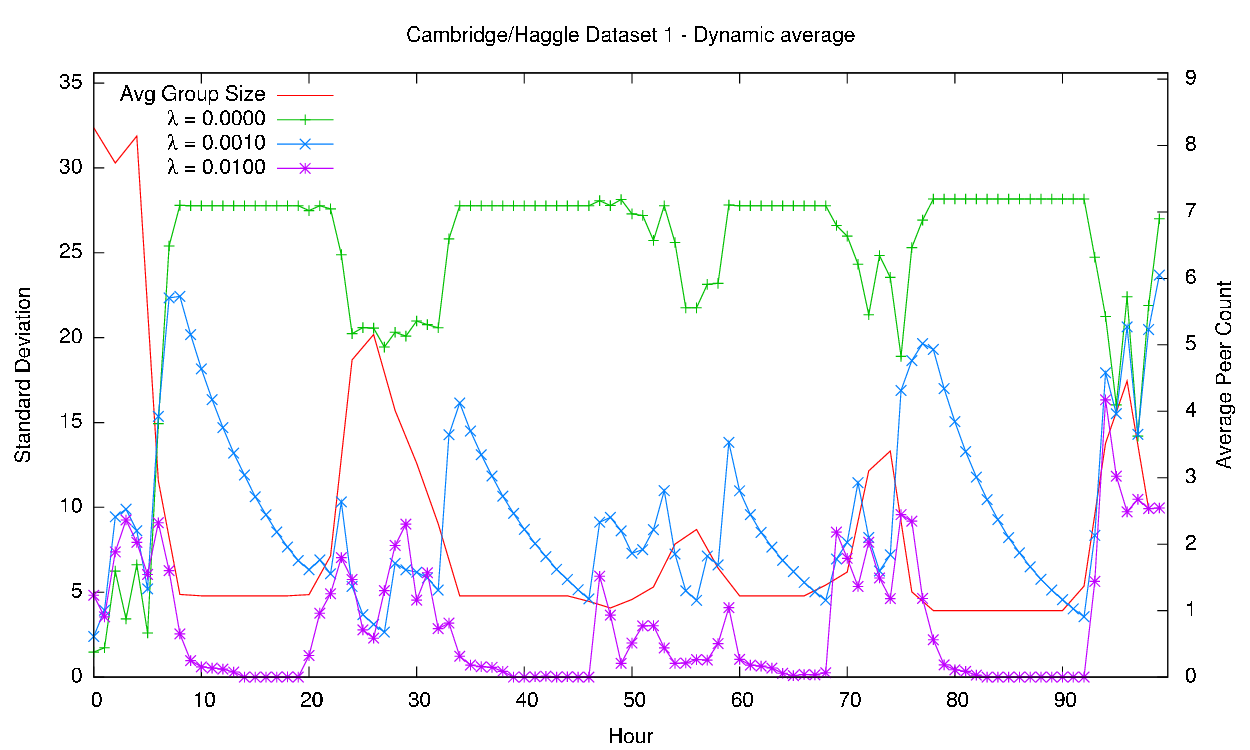} & \includegraphics[width=3.5in]{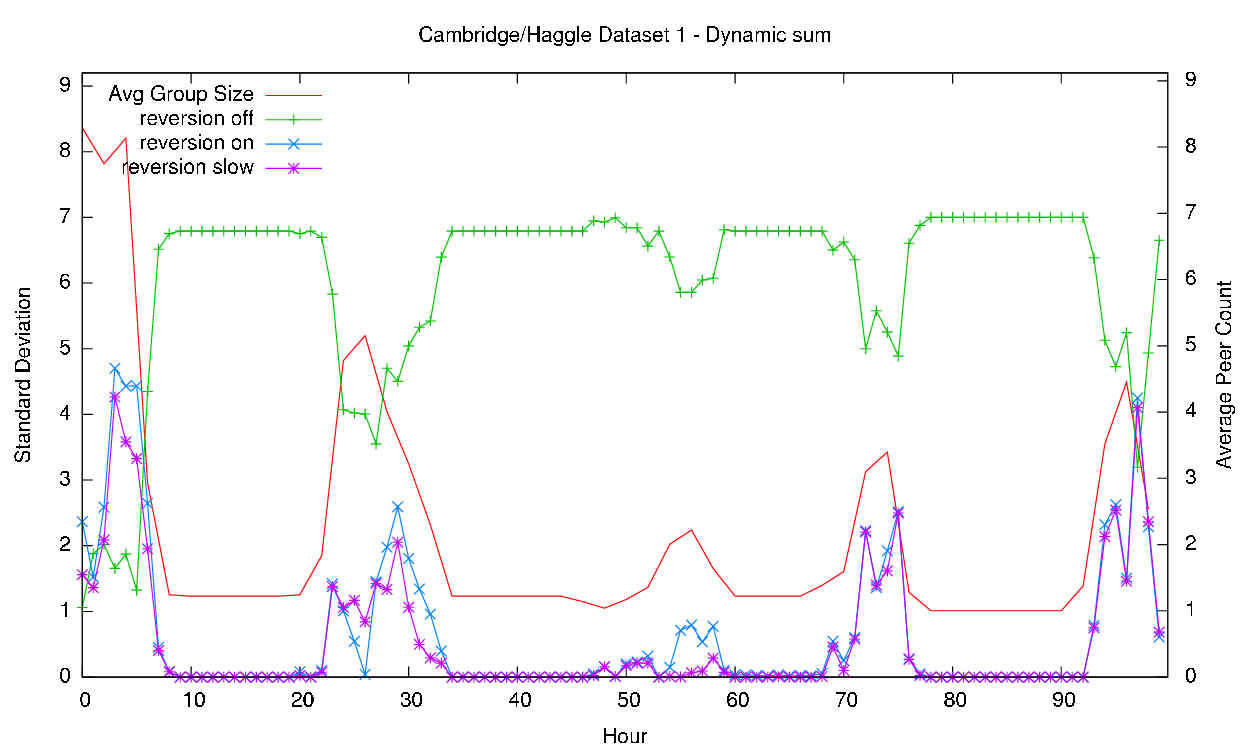} \\
\includegraphics[width=3.5in]{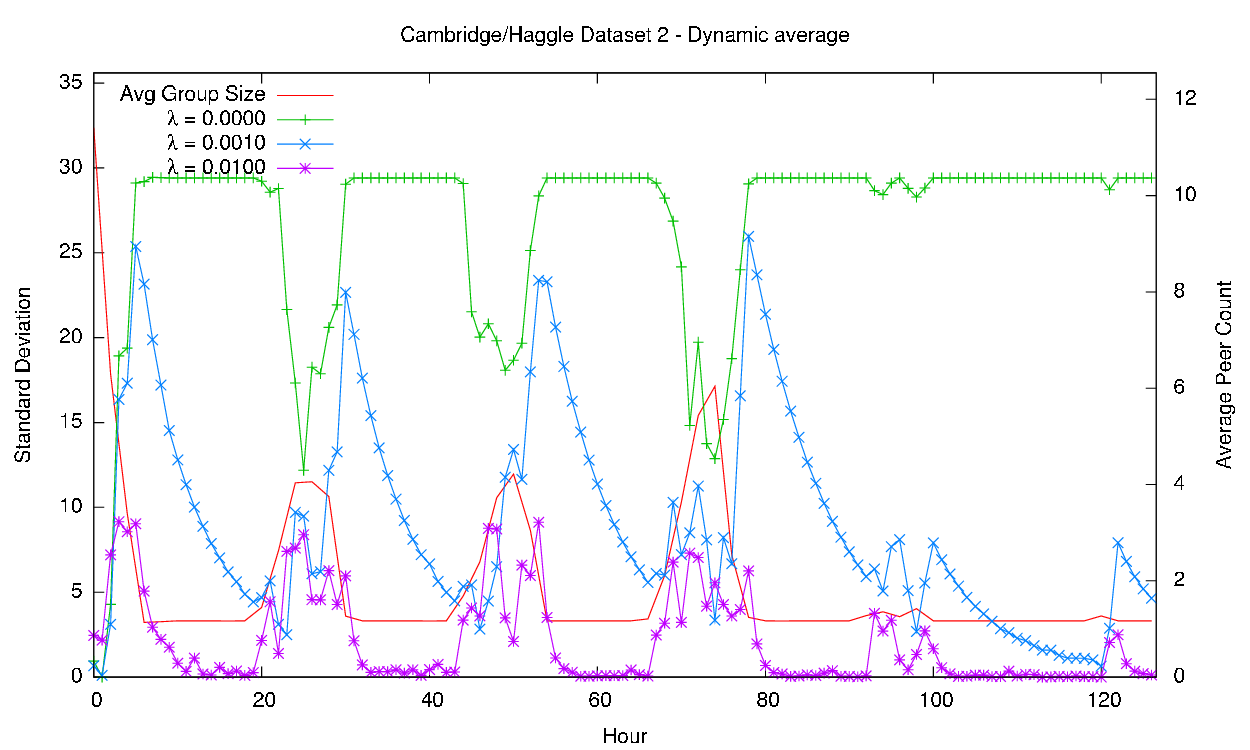} & \includegraphics[width=3.5in]{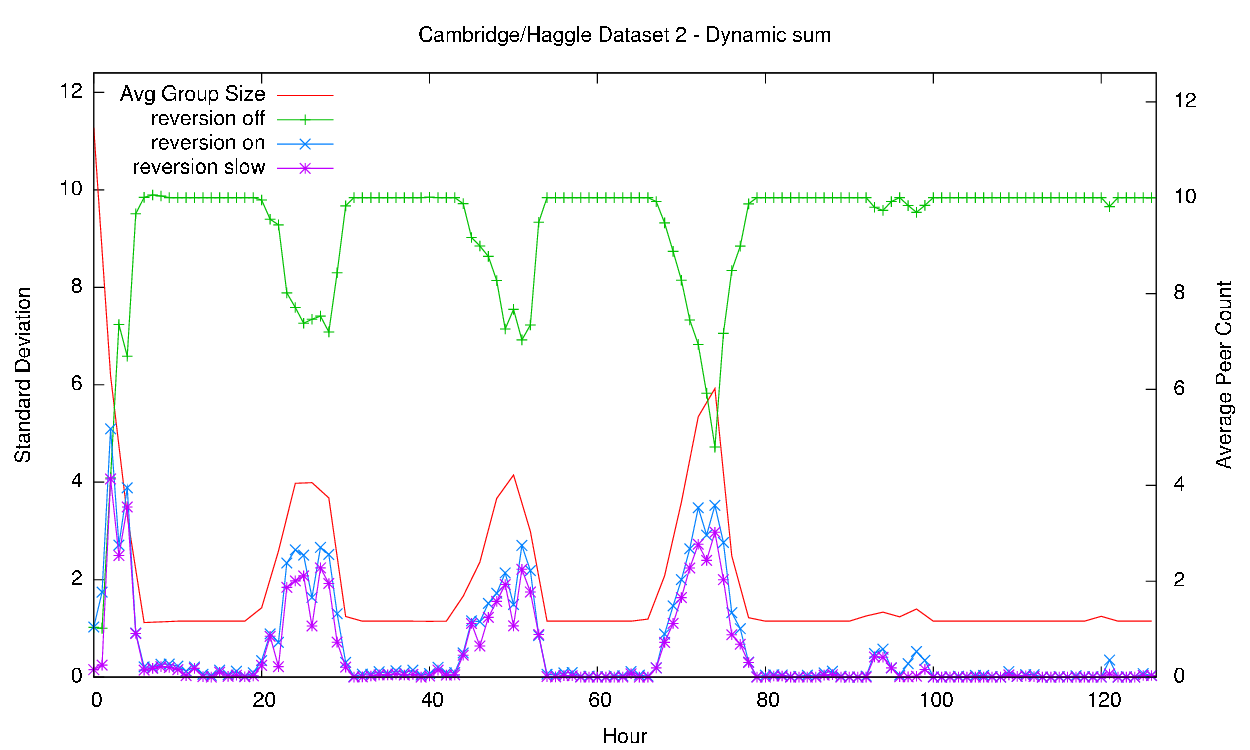} \\
\includegraphics[width=3.5in]{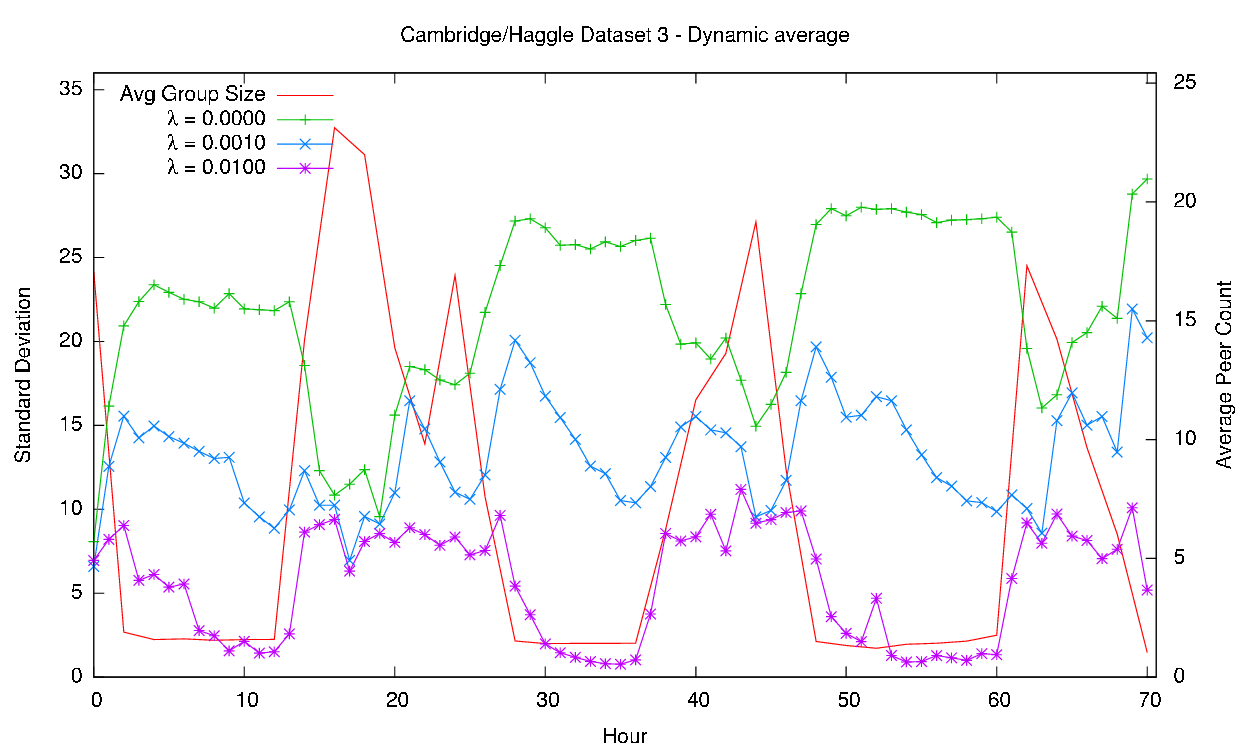} & \includegraphics[width=3.5in]{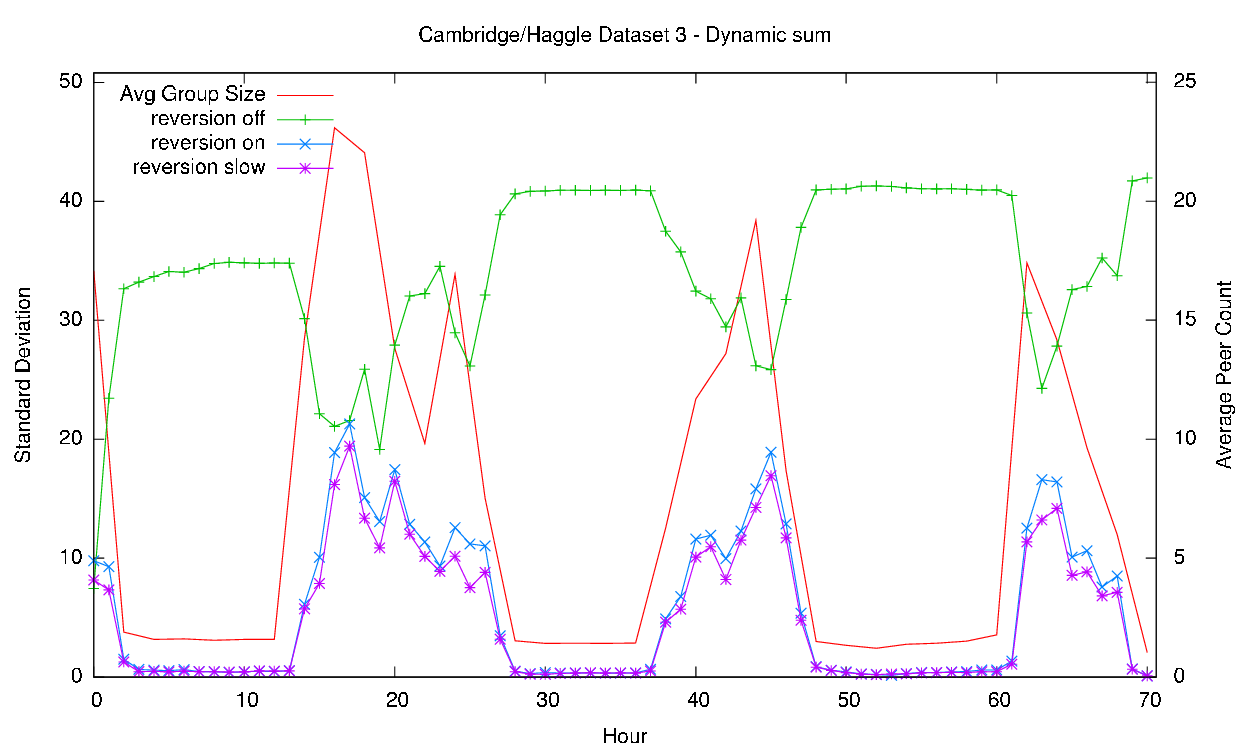}
\end{tabular}
\caption{\textbf{Behavior of dynamic averaging and summation on the Cambridge/Haggle datasets\ \cite{cambridge-haggle-2006-09-15}.} The standard deviation from the correct estimate of running group average and size as a function of time.  Gossip was performed once every 30 seconds.  The data sets include traces of 9, 12, and 41 wireless devices, worn by students over the course of several days.  The devices recorded their adjacency matrix as a function of time.  For reference, actual running group size is also presented on the same graphs.}
\label{fig:average:haggle}
\label{fig:sum:haggle}
\end{center}
\end{figure*}

\subsection{Push-Sum-Revert}
Push-Sum-Revert maintains a running estimate of values stored across the entire network, eliminating the influence of nodes no longer in the system.  The reversion factor $\lambda$ controls how rapidly this influence is removed.  A reversion factor of zero corresponds to the Push-Pull variant of traditional Push-Sum.  In all Push-Sum-Revert tests, the expected average is $50$, and standard deviation should be considered with respect to this value.

In order to properly analyze the protocol, we consider two different failure modes.  By the law of large numbers, random host failures do not impact the average over the long term, nor will they affect the mass in the system.  Traditional push-sum is sufficient to maintain a running estimate if there is no correlation between node failures and their values.  In the interest of completeness, we include an analysis of Push-Sum-Revert under these conditions.  As Figure \ref{fig:average:uncorrelated} shows, massive uncorrelated node failures (20 steps into the test) have no direct adverse effects on any instance of Push-Sum-Revert.  

However, host failures that are correlated with values stored at those hosts will alter the average without altering the average mass in the system.  We simulate this by failing the highest valued half of the participating nodes.  This lowers the expected average to $25$.

To recover from the error introduced by such failures, Push-Sum-Revert forces the system towards its initial value by a factor of $\lambda$ at each iteration.  As Figure \ref{fig:average:correlated}a shows, higher values of $\lambda$ result in faster convergence but result in greater error once the system has converged.  This error can be reduced by forcing nodes to export their entire mass at each iteration.  

Figure \ref{fig:average:correlated}b demonstrates the benefits of the full-transfer optimization.  At each iteration, all hosts split their mass into 4 parcels and send each parcel to a different peer.  The increased variance in per-round estimate is counteracted by averaging over the mass received during the last 3 iterations (iterations during which the host received no mass are skipped).  With a reversion constant of $\lambda=0.5$ the standard deviation of the protocol drops to under 10 rounds with a standard deviation of $2.13$ ($8.53\%$ of the actual average).  Lowering the constant to $\lambda=0.1$ increases convergence time to 35 rounds, but lowers the standard deviation to $0.694$ ($2.77\%$ of the actual average).  

By comparison, the traditional protocol takes 10 rounds to converge on a network of this size.  Consequently, a protocol based on periodic resets would take at least 10 gossip iterations to compute this estimate.  This figure does not account for the possibility of larger networks, different value distributions, or lack of clock synchronization.  Push-Sum-Revert reacts to all of these situations dynamically.

In low connectivity situations, the error introduced by reversion constants grows more rapidly.  The protocol continues to outperform traditional Push-Sum.  Figure \ref{fig:average:haggle} shows the behavior of Push-Sum-Revert on the Cambridge/Haggle datasets.  Particularly in cases where hosts frequently form small groups (as in dataset 1), Push-Sum-Revert significantly outperforms traditional Push-Sum.  Note that because of the small number of participants, the variance between subset averages was sufficient to cause visible errors.  Just to emphasize the point, node departures are {\em not} correlated with their values in this figure.

\subsection{Dynamic Sketch Counting}

The non-uniqueness of host identifiers in counting sketches makes it impossible for hosts to independently determine whether they are the sole source of a bit.  Consequently, unlike distributed aggregation, traditional sketch counting is affected by any failure model that removes hosts.  The range cutoff used by Count-Sketch-Reset limits how long a bit no longer sourced remains in the system.  Unlike Push-Sum-Count's $\lambda$, the effect of raising the cutoff drops steeply after a certain point.  As shown in Figure \ref{fig:pushsumrevert}, under uniform gossip the cutoff for bit k is approximated by $7 + \frac{k}{4}$.  The following experiments use 64 buckets for an expected error of 9.7\%\ \cite{Flajolet}.

Figure \ref{fig:sum} shows the behavior of Count-Sketch-Reset in a uniform gossip environment.  The algorithm introduces a small amount of error at several nodes, but reverts to its original state within 10 rounds of a massive node failure.

Figure \ref{fig:sum:haggle} addresses real world environments.  Because of the small number of nodes participating in the traces (relative to the size of our other environments), each node acquires 100 identifiers and adjusts its estimate of the network size accordingly.  This increases the average value of $R(A)$ without reducing the gossip propagation time.  Consequently, the effective reversion factor is higher.  The protocol estimate's standard deviation remains within half of the correct value for the duration of both traces 1 and 2, performing significantly better than the flat estimate generated by the basic protocol.

\section{Related Work}
\label{sec:relatedwork}
This work builds on weighted averaging as developed by Kempe et al\ \cite{Kempe-Gossip}, as well as a variety of work based on counting sketches \ \cite{Flajolet,Considine,Alon}.  Both approaches converge on the desired aggregate by allowing randomly selected host pairs to converge individually.  Because piecewise convergence is achieved through zero-net-change exchanges, every interaction causes individual host estimates to converge on the global aggregate.  We extend these protocols to maintain a running estimate of the aggregate, responding to changes in network configuration by continuing to converge on the correct value.  

Jelasity and Montresor\ \cite{Jelasity} describe a technique similar to Kempe et al's Push-Sum.  They further extend this technique to make use of epochs in order to maintain a running estimate of the sum and average.  Though more accurate than introducing local error, the use of epochs requires inter-host coordination.  As already noted, this coordination may be infeasible in certain situations.

Kostoulas et al\ \cite{Kostoulas} present two algorithms for estimating the number of hosts in a dynamic cluster.  The first of these, Hops Sampling, measures the average gossip distance between every host in the network and a predefined leader host. In a uniform environment, this value is $\Theta(log(n))$ in the size of the cluster.  In the second protocol, Interval Density, a leader host passively listens to network traffic and records the unique randomly assigned identifiers of observed hosts.  Sampling only a subset of the identifier space limits the size of this list, while using uniformly distributed identifiers ensures that the sample is applicable to the entire space.  

While both approaches are able to compute the network size using less bandwidth than Count-Sketch-Reset, both also rely on a central coordinator host.  In highly distributed environments such as a wireless network, deployment of a coordinator can be logistically infeasible.  Even in situations when it is feasible to deploy a single coordinator, the decentralized approach taken by Push-Sum-Revert and Count-Sketch-Reset removes the single point of failure presented by the coordinator.

More generally, gossip protocols were first introduced by Demers et al\ \cite{Demers} for use in distributed database maintenance, where they were observed to have exponential convergence times. They note that such protocols break down into three forms: push (hosts randomly select peers to send gossip to), pull (hosts randomly select peers to accept gossip from), and push/pull (hosts randomly select peers to exchange gossip with).  They note that the use of pull-based gossip is preferable in cases where gossip is already widespread, a fact used by Karp et al\ \cite{Karp} to refine Demers et al's $O(n\log(n))$ bound down to $O(n\log\log(n))$ for the case of push/pull algorithms.  We exploit this by incorporating push/pull into our protocol simulations.

Other work on in-network aggregation involves the imposition of structure onto the network.  TAG\ \cite{Madden}, Mobile Agents\ \cite{Pinhero} and SPIN\ \cite{Heinzelman} attempt to limit the bandwidth used in data-dissemination protocols by performing aggregation inside the network.  All three flood small user requests for data through the entire network and then use the flood path to build a spanning tree.  Data is then passed up the spanning tree and aggregated where possible.  A slightly more general approach is taken by Directed Diffusion\ \cite{Intanagonwiwat}, where the spanning tree adapts to requests issued by different parts of the network.  However, despite this adaptability, a relatively stable network is still assumed.  Push-Sum-Revert and Count-Sketch-Reset introduce a small, bounded error into the results, but are able to operate in an entirely unstructured dynamic network environment.

\section{Conclusion}

We have presented a general class of distributed aggregation protocols: the dynamic distributed aggregation protocol. Protocols in this class are fundamental to the design of decentralized wireless infrastructures. We have proposed several variant algorithms that may be used to maintain the average, count, and sum aggregates in such dynamic environments.  These three techniques demonstrate how small, directed errors may be used to maintain estimates of aggregates over volatile networks.

\bibliographystyle{IEEEtran}
\bibliography{distributed-agg.bib}
\end{document}